\documentclass[aps,prd,twocolumn,showpacs,nofootinbib,amsmath,amssymb,amsfonts,showkeys]{revtex4-1}

\usepackage{bm}
\usepackage{mathtext}
\usepackage[T1,T2A]{fontenc}
\usepackage[utf8]{inputenc}
\usepackage[english]{babel}
\usepackage[dvips]{graphicx}
\usepackage{xcolor}

\begin{document} 

\title{Signal in the hyperfine structure line of the ground state \\ of atomic hydrogen from the Dark Ages as a cosmological test}

\author{B. Novosyadlyj$^{1,2}$, Yu. Kulinich$^1$, O. Konovalenko$^3$}
\affiliation{$^1$Astronomical Observatory of Ivan Franko National University of Lviv, \\
8 Kyryla and Methodia str., Lviv, 79005, Ukraine }
\affiliation{\it $^2$International Center for Future Sciences of Jilin University, \\
2699 Qianjin str., Changchun, 130012, P.R. China}
\affiliation{\it $^3$Institute of Radio Astronomy of NASU, 4 Mystetstv str., Kharkiv, 61002, Ukraine}

\date{\selectlanguage{english}\today}
 
\begin{abstract}
We analyze the formation of the signal in the 21 cm hydrogen line of the Dark Ages ($30\le z\le300$) in different cosmological models and discuss the possibility of its detection by decameter-wavelength radio telescopes. To study the dependence of the intensity and profile of the line on the values of cosmological parameters and physical conditions in the intergalactic medium, the evolution of the global (averaged over the sky) differential brightness temperature in this line was calculated in standard and non-standard cosmological models with different parameters. The standard $\Lambda$CDM model with cosmological parameters predicts a value of the differential brightness temperature in the center of the absorption line $\delta T_{br}\approx-35$ mK at $z\approx87$. The frequency of the line at the absorption maximum is 16 MHz, the effective width of the line is $\approx$25 MHz. The depth of the line is moderately sensitive to $\Omega_b$ and $H_0$, weakly sensitive to $\Omega_{dm}$ and insensitive to other parameters of the standard $\Lambda$CDM model. However, the line is very sensitive to additional mechanisms of heating or cooling of baryonic matter during the Dark Ages, so it can be an effective test of non-standard cosmological models. In models with decaying and self-annihilating dark matter, as well as with an initial global stochastic magnetic field, the temperature of baryonic matter in this period is higher, the higher the density of these dark matter components and the magnetic field strength. The absorption line becomes shallower, disappears, and turns into emission at values of the component parameters lower than the upper bounds on them, which follow from observational data. The estimations show that such spectral features can be detected by decameter-wavelength radio telescopes in the near future.
\end{abstract}
\pacs{95.36.+x, 98.80.-k}
\keywords{cosmological Dark Ages, 21 cm hydrogen line, self-annihilating dark matter, decaying dark matter, primordial magnetic fields }
\maketitle
 

\section{Introduction}

The discovery of two galaxies with intense star formation at redshifts 12.63 (JADES-GS-z12-0) and 13.20 (JADES-GS-z13-0) \cite{Robertson2023,Curtis-Lake2023} has renewed interest in the early epochs of the evolution of the Universe, when the first cosmic objects began to form. Their masses are $2.5\cdot10^8$ М$_\odot$ and $6.3\cdot10^7$ М$_\odot$, respectively. The existence of galaxies of such masses and such concentration in the young Universe (340 and 320 million years after the Big Bang) is a challenge for modern cosmology. Another topical issue is the discrepancy at the level  $\geq4\sigma$ between the the Hubble constant values determined from type Ia supernovae in the nearby (late) Universe ($z\le3$, $H_0=74.0\pm1.4$ km/s$\cdot$Mpc) and from the power spectrum of temperature and polarization fluctuations of the cosmic microwave background (CMB) radiation, which was formed at the early Universe ($z>1000$, $H_0=67.4\pm0.5$ km/s$\cdot$Mpc) \cite{Verde2019,Riess2022}. These problems motivate us to search for new cosmological tests, in particular, at intermediate redshifts $z\sim10-300$, in the epoch of the Dark Ages and Cosmic Dawn, when there were no stars or galaxies yet. What signals can we expect from those times?

An important information channel about the state of baryonic matter in this period is line of the hyperfine structure of the ground state of hydrogen $\lambda_0$=21 cm (see reviews \cite{Barkana2001,Fan2006,Furlanetto2006,Bromm2011,Pritchard2012,Natarajan2014,Shimabukuro2022,Minoda2023}). This is the wavelength of the radiation in the rest frame, its frequency in vacuum is $\nu_0$=1420 MHz. For an terrestrial observer, this line is shifted to longer wavelengths $\lambda_{obs}=21(1+z)$ cm ($\nu_{obs}=1420/(1+z)$ MHz) due to the cosmological expansion of the Universe, where $z$  is the redshift of the region where the line is formed. The earliest signals of the formation of the halos, or the first galaxies, in the Dark Ages can be obtained in this spectral line  also \cite{Iliev2002,Iliev2003,Furlanetto2006a,Shapiro2006,Kuhlen2006,Novosyadlyj2020}. The physical conditions in the pre-reionization epoch, the state of excitation and ionization of hydrogen, have been analyzed in different cosmological models and scenarios of the formation of the first light sources. Known spectral features include broad absorption lines, shifted to $\sim10-20$ MHz at $z\sim90-80$ and $\sim70-130$ MHz при $z\sim20-10$, as well as a weak emission line before full reionization. The second absorption line is caused by the Wouthuysen–Field effect \cite{Wouthuysen1952,Field1958,Field1959} and is determined mainly by the spectral energy distribution of the first light sources \cite{Pritchard2010,Mirocha2013,Mirocha2015,Cohen2017,Monsalve2017,Monsalve2018,Monsalve2019,Novosyadlyj2023}.
In recent years, several projects have been implemented to detect the 21 cm absorption line, which is formed in the redshift range $10-20$ (observed radio wavelength $\sim$2--4 m, frequency $\sim68-130$ MHz). The only detection of this line in the Experiment to Detect the Global Epoch of Reionization Signature (EDGES) experiment \cite{Bowman2018} indicates an unusual profile shape and unexpectedly large depth with a center at 78 MHz, which is $\sim$3--4 times deeper than expected in the standard $\Lambda$CDM model. Explanations go beyond the standard cosmology, including additional mechanisms for baryons cooling, excess radio background at high redshifts, viscous dark energy, etc. \cite{Barkana2018,Ewall-Wice2018,Halder2022}. Another explanation \cite{Hills2018} lies in the difficulty of removing foreground noise, removing synchrotron radiation from the interstellar plasma of our galaxy, radiation from the Earth's ionosphere, and man-made radio emission, which are several orders of magnitude higher in intensity than the useful extragalactic signal in this wavelength range. Recent similar measurements in the SARAS~3 experiment (Shaped Antenna measurement of the background Radio Spectrum 3) \cite{Singh2022} confirm the latter assumption: the registered spectrum does not show any features found in EDGES, with a confidence level of 95.3\%. However, SARAS~3 \cite{Singh2022} did not detect any spectral features in the frequency range 55–85 MHz, so both theoretical predictions and measurements of the 21 cm signal from the early Universe are still relevant.

\begin{figure*}[htb]
\includegraphics[width=0.495\textwidth]{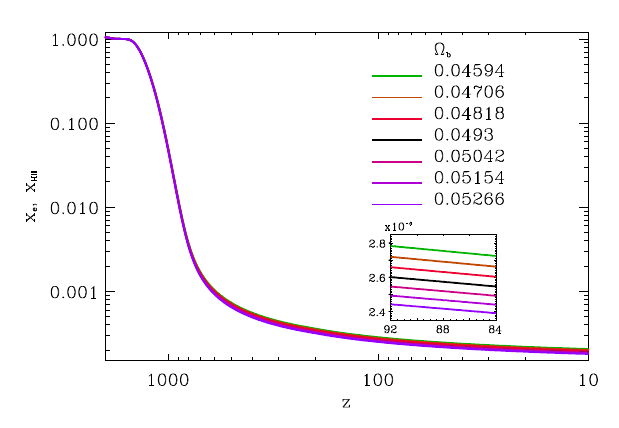}
\includegraphics[width=0.495\textwidth]{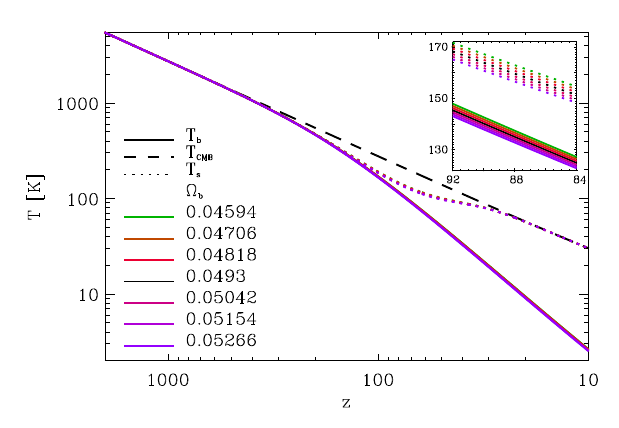} 
\caption{Dependence of the fraction of ionized hydrogen (left) and temperature (right) on redshift in the epochs of cosmological recombination and the Dark Ages in the standard $\Lambda$CDM model. Solid lines show the temperature of baryonic matter in the models with $\Omega_b=0.0493\pm1\sigma,\,\pm2\sigma,\,\pm3\sigma$, dotted lines show the corresponding spin temperatures, and dashed lines show the CMB temperature. The insets show the dependence of the ionized hydrogen fraction, gas temperature, and spin temperature on the redshift in the region of formation of the HI 21 cm absorption line.}
\label{xeT_Omb}
\end{figure*} 
\begin{figure}
\includegraphics[width=0.5\textwidth]{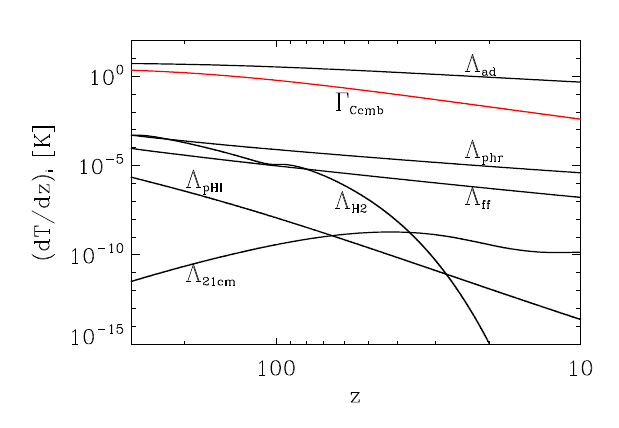} 
\caption{Heating/cooling functions in the intergalactic medium of the Dark Ages in the standard $\Lambda$CDM model.}
\label{hcf}
\end{figure} 

Absorption line of 21 cm, formed in an earlier epoch at $z\sim100$, in the Dark Ages, is less studied in terms of its informativeness and detection capabilities. For a terrestrial observer, it is located in the decameter wavelength range $\lambda_{obs}\sim20$ m ($\nu_{obs}\sim15$ MHz), which is even more difficult for detecting extragalactic signals from the Earth's surface. To eliminate terrestrial interference (of both natural and man-made origin), scientific teams from different countries are proposing projects for decameter wavelength radio telescopes located on the far side of the Moon \cite{Bentum2020,Rapetti2023,Plice2017,Goel2022,Shkuratov2019}, including the team of the Institute of Radio Astronomy of the NAS of Ukraine. Two of them \cite{Bentum2020,Rapetti2023} are already in the implementation stage.

In this paper, we analyze the sensitivity of the position and depth of the global (averaged over the entire sky) 21 cm absorption line of neutral hydrogen to the cosmological parameters of the standard $\Lambda$CDM model, as well as the heating/cooling of the baryonic component in non-standard cosmological models with decaying and self-annihilating dark matter, a primordial magnetic field, and additional cooling. We consider the standard cosmological model to be $\Lambda$CDM with cosmological parameters from \cite{Planck2020a,Planck2020b}: $H_0=67.36\pm0.54$ km/s$\cdot$Mpc (Hubble constant), $\Omega_b=0.0493\pm0.00112$ (density parameter\footnote{Density parameter is the mean density of the component in the current epoch in units of critical: $\Omega_i\equiv\rho^0_i/\rho^0_{cr}$ where $\rho^0_{cr}\equiv 3H_0^2/8\pi G$} of baryonic matter), $\Omega_{dm}=0.266\pm0.0084$ (density parameter of dark matter), $\Omega_r=2.49\cdot10^{-5}$ (density parameter of relativistic component (CMB and neutrinos)), $\Omega_{\Lambda}=0.6847\pm0.0073$ and $\Omega_K=0$ (dimensionless\footnote{$\Omega_{\Lambda}\equiv c^2\Lambda/3H^2_0$, $\Omega_{K}\equiv -c^2K/H^2_0$} cosmological constant $\Lambda$ and curvature of 3-space $K$, respectively). 
\begin{figure*}[htb]
\includegraphics[width=0.495\textwidth]{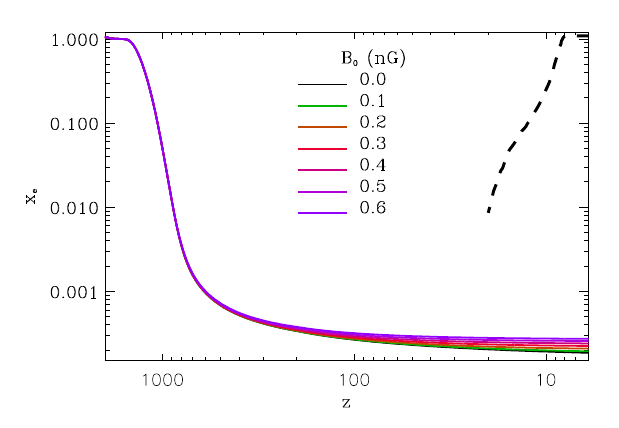}
\includegraphics[width=0.495\textwidth]{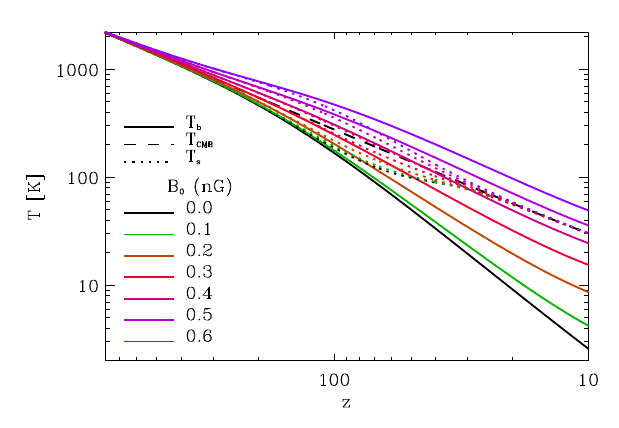}
\caption{Dependence of ionized hydrogen fraction (left) and baryonic matter temperature (right) on redshift int he  models with primordial stochastic magnetic fields with different rms values of induction $B_0$, which increase for lines from bottom to top. The dashed line in the left panel  is the $2\sigma$ upper limit from the CMB radiation polarization data \citep{Planck2020a}.}
\label{xT_pmf}
\end{figure*}

\begin{figure}[htb]
\includegraphics[width=0.5\textwidth]{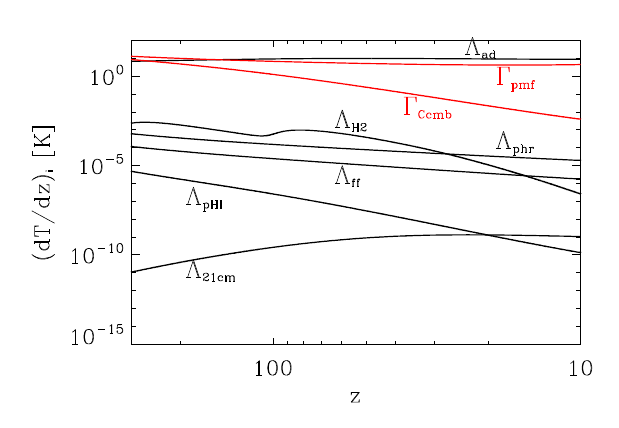} 
\caption{Heating/cooling functions in the model with $B_0$=0.6 nG.}
\label{hcf_pmf}
\end{figure}

The plan of the paper is as follows. In Chapter II, we describe the state of the intergalactic medium and, in particular, the ionization and excitation of neutral hydrogen in different cosmological models during the Dark Ages, from cosmological recombination to the beginning of reionization in the Cosmic Dawn era. In Chapter III, we analyze the dependence of the position and depth of the 21 cm absorption line in the Dark Ages on cosmological parameters in the standard $\Lambda$CDM model and additional heating and cooling of the baryonic component in non-standard cosmological models. In Chapter IV, we analyze the capabilities and prospects for detecting this line with existing and planned decameter-wavelength radio telescopes. The results are summarized in the Conclusions. 

\section{The State of the Intergalactic Medium in the Dark Ages}

For the calculation of the emission of neutral hydrogen in the 21 cm line in the Dark Ages epoch, it is necessary to calculate the ionization and thermal history of the gas, as well as the population of levels of the hyperfine structure of atomic hydrogen. All of them depend on the parameters of the cosmological model, the ionization mechanisms, heating and cooling of the baryonic gas. In this section, we will consider the ionization and thermal history of the gas in the standard $\Lambda$CDM model, in the model with primordial magnetic field, with decaying and self-annihilating dark matter, as well as in the model with additional cooling. This is far from all the possibilities of expanding the standard $\Lambda$CDM model, but those that demonstrate the prospects for setting up observational studies in this direction.

\subsection{Standard $\Lambda$CDM model}

After the epoch of cosmological recombination and before the epoch of reionization, almost all of the hydrogen in the Universe, > 99\%, was in a neutral state. The residual ionization of the Universe is a consequence of its expansion. The fraction of the ionized fraction $x_{HII}$ is less than  1\% at $30<z<700$ and depends on the density, the mechanisms of heating/cooling of the gas, and the expansion rate of the Universe. For a homogeneous isotropic Universe that is expanding, the temperature of the gas and the ionization of atoms are calculated by integrating a system of differential equations of energy balance and ionization-recombination kinetics of atoms\footnote{The first molecules make up a small fraction of all particles (see \cite{Novosyadlyj2017,Kulinich2020,Novosyadlyj2020a,Novosyadlyj2022}) and are not discussed in this work}. In the standard $\Lambda$CDM  model, the main mechanism of heating of the gas in this period is Compton scattering of the CMB radiation on free electrons, and for cooling is adiabatic expansion. All other mechanisms that occur in astrophysical plasma under the conditions of the Dark Ages are weaker by 3 or more orders of magnitude (see Fig. 4 and Appendix in \cite{Novosyadlyj2023}). The temperature of the plasma and the ionization of hydrogen and helium during cosmological recombination and in the epoch of the Dark Ages up to $z=200$ are calculated on the basis of an effective 3-level atomic model implemented in the publicly available program RecFast\footnote{http://www.astro.ubc.ca/people/scott/recfast.html} \citep{Seager1999,Seager2000}, and for $z<200$ the system of equations
\begin{eqnarray}
&&\hskip-0.5cm -\frac{3}{2}n_{tot}k_B(1+z)H\frac{dT_b}{dz} = \Gamma_{C_{cmb}}-\Lambda_{ad}- \Lambda_{ff}-    \nonumber \\
&&\hskip1cm-\Lambda_{phr}-\Lambda_{21cm} -\Lambda_{H_2} + \Gamma_{nSM} - \Lambda_{nSM'},\label{Tb}\\
&&\hskip-0.5cm -(1+z)H\frac{dx_{HII}}{dz}=R_{HI}x_{HI}+C^i_{HI} n_i x_{HI}-  \nonumber \\
&&\hskip4.5cm-\alpha_{HII}x_{HII}x_en_H, \label{kes}\\
&&\hskip-0.5cm -(1+z)H\frac{dx_{HeII}}{dz}=R_{HeI}x_{HeI}+C^i_{HeI} n_i x_{HeI}- \nonumber \\
&&\hskip4.3cm-\alpha_{HeII}x_{HeII}x_en_H, \label{kes2}\\
&&\hskip-0.5cm H=H_0\sqrt{\Omega_m(1+z)^3+\Omega_K(1+z)^2+\Omega_{\Lambda}}, \label{H}
\end{eqnarray}
where $T_b$ is the temperature of the baryonic gas, $n_{tot}\equiv n_{HI}+n_{HII}+n_{e}+n_{HeI}+n_{HeII}=n^0_{tot}(1+z)^3$ is the concentration of particles of all types, $n_H\equiv n_{HI}+n_{HII}=n^0_H(1+z)^3$ is the concentration of hydrogen, $k_B$ is the Boltzmann constant, $\Gamma_{C_{cmb}}$ is the heating function due to Compton scattering of CMB quanta on free electrons \cite{Seager1999,Seager2000},    
$\Lambda_{ad}$ is the adiabatic cooling function due to the expansion of the Universe \cite{Seager1999}, $\Lambda_{ff}$ is the cooling function due to free-free transitions of electrons \citep{Shapiro1987}, $\Lambda_{phr}$ is the cooling function due to photorecombination \citep{Anninos1997}, $\Lambda_{21cm}$ is the cooling function due to emission in the 21 cm line \citep{Seager1999}, $\Lambda_{H_2}$ is the cooling function due to emission in rotational-vibrational transitions of the $H_2$ molecule \citep{Seager1999}, $\Gamma_{nSM}$ and $\Lambda_{nSM'}$ are the heating and cooling functions in non-standard models (see next subsections), $R_{HI},\,R_{HeI}$ are the photoionization rates of hydrogen and helium atoms, $C^i_{HI},\,C^i_{HeI}$ are the shock ionization coefficients of the $i$-th type of particles, $\alpha_{HII},\,\alpha_{HeII}$ are the recombination coefficients of hydrogen and helium ions, $\Omega_m=\Omega_b+\Omega_{dm}$, $H(z)$ is the expansion rate of the Universe.

Fig. \ref{xeT_Omb} shows the calculations of the ionization and thermal history of the medium from the epoch of cosmological recombination of hydrogen ($z=2000$) to the end of the Dark Ages epoch ($z=10$). The scatter of the $\Omega_b$ value within 3$\sigma$ ($\pm$6.8\% from the most optimal value of 0.0493) leads to a 7.2\% spread of $x_e$ around the value of $1.918\cdot10^{-3}$ and a 2.7\% spread of $T_b$ around the value of 2.563 K at $z=10$. The main heating mechanism for the gas in this period is Compton scattering of CMB quanta on free electrons, and for cooling is adiabatic expansion. All other mechanisms that occur in astrophysical plasma in the Dark Ages conditions in the standard $\Lambda$CDM model are 3 or more orders of magnitude weaker, as illustrated well by Fig. 2 However, in non-standard cosmological models, with additional mechanisms of ionization, heating, or cooling, the thermal and ionization history of the plasma of the Dark Ages epoch may be different. 

\subsection{Model with primordial magnetic fields}

The inclusion of the primordial magnetic fields (PMFs) in the cosmological scenarios of the large-scale structure formation of the Universe, on one hand, is due to hypothetical mechanisms for their generation in the early Universe (see review \cite{Subramanian2016}) and, on the other hand, to experimental evidence for the existence of extragalactic magnetic fields, which follow from the data of observations of blazar radiation in the GeV and TeV energy ranges by the Fermi Space Telescope, the H.E.S.S. (High Energy Stereoscopic System) stereoscopic system of $\gamma$-ray receivers in Namibia, and the ARGO-YBJ detector array in Tibet \cite{Neronov2010,Takahashi2013}. The latter indicate a lower limit on the value of the induction of such fields on scales of several megaparsecs: $B_0>10^{-20}$ Gs. The upper limit was obtained from the studies of the influence of the stochastic background of PMFs on the anisotropy of the temperature and polarization of the CMB in the Planck experiment \cite{Planck2016}: $B_0<9\cdot10^{-10}$ Gs.

Magnetic fields decay, which leads to the heating of baryonic matter, affecting the thermal and ionization history of the Universe in the Dark Ages. Two mechanisms of heating of baryonic matter are considered: due to the damping of the turbulent component of the magnetic field and ambipolar diffusion. The dependence of the heating function for them on redshift, taking into account the results in \cite{Sethi2005,Chluba2015,Mack2002,Minoda2019} can be represented as follows: 
\begin{eqnarray*}
&&\Gamma_{mfdt}(z\ge z_{rec}) = 1.5\rho_{mf}H(z)\frac{m}{a}[f_D(z)]^{n_B+3}\times \nonumber \\
&&\hskip2.5cm\exp\left\{-\frac{(z-z_{rek})^2}{5000}\right\}\left(\frac{1+z_{rek}}{1+z}\right)^4, \nonumber \\
&&\Gamma_{mfdt}(z<z_{rec}) = 1.5\rho_{mf}H(z)[f_D(z)]^{n_B+3}\times \nonumber \\
&&\hskip2.5cm\frac{ma^m}{(a+1.5\ln((1+z_{rek})/(1+z)))^{m+1}}, \\ 
&&\Gamma_{mfad}(z) = \frac{1-x_{HII}}{g(T_b)x_{HII}}[f_D(z)]^{2n_B+8}\times \nonumber \\
&&\hskip4.5cm\left[\frac{(1+z)k_D}{3.086\cdot10^{22}}\frac{\rho_{mf}}{\rho_b}\right]^2f_L, \label{mfad}
\end{eqnarray*}
where $[\Gamma]$=erg/cm$^3$/s, $z_{rek}$ and $t_{rec}$ are redshift and time of cosmological recombination accordingly, $\rho_{mf} =3.98\cdot10^{-20}\left(B_0/{\text{nGs}}\right)^2(1+z)^4$~erg/cm$^{3}$ is energy density of the PMF, $n_B=-2.9$ is the spectral index of the PMF power spectrum, $a = \ln(1+t_\mathrm{d}/t_\mathrm{rec})$, $m \equiv 2(n_B+3)/(n_B+5)$, $t_\mathrm{d}/t_\mathrm{rec}=14.8/(B_0k_D)$, $k_D=(2.89\cdot10^4 h)^{1/(n_B+5)}B_\lambda^{-2/(n_B+5)}k_\lambda^{(n_B+3)/(n_B+5)}$ Mpc$^{-1}$, $\lambda = 1$ Mpc, $B_\lambda = B_{1\,\mathrm{Mpc}} =B_0$, $k_\lambda = k_{1\,\mathrm{Mpc}} = 2\pi$ Mpc$^{-1}$,  $x_{HII} = n_{HII}/n_{H}$, $f_L=0.8313(n_B+3)^{1.105}(1.0-0.0102(n_B+3))$, $g(T_b) = 1.95\cdot10^{14} T_b^{0.375}$~cm$^3$/s/g, $\rho_b = \rho_{cr}^{(0)}\Omega_b(1+z)^3$, $k_D=286.91(B_0/\text{nGs})^{-1}$~Mpc$^{-1}$. The factor $f_D(z)^{n_B+3}$ describes the evolution of the cutoff scale of the primordial magnetic field spectrum (see \cite{Minoda2019}), which we approximate for $z\le1178$ and fixed values of $n_B$ and $k_D$ by a 4th-order polynomial
$[f_D(z)]^{n_B+3} \simeq 0.6897525+0.2944149\cdot10^{-3}z-0.3805730\cdot10^{-6}z^2+0.2259742\cdot10^{-9}z^3+0.6354026\cdot10^{-13}z^4$. For $z\ge1178$ $f^{n_B+3}(z) \equiv 1$.

Results of calculations for the thermal and ionization state of baryonic matter in the models with different root-mean-square values of the field induction $B_0=$0.1, 0.2, ... 0.6 nGs are shown in Fig. \ref{xT_pmf}. As we can see, PMF with $B_0$ smaller than upper observational limits noticeably increase the temperature of baryonic matter at $z<300$. If $B_0\gtrsim0.4$ nGs, then the temperature $T_b$ throughout the Dark Ages epoch is greater than the temperature of the CMB. The fraction of ionized hydrogen in this epoch slightly increases, remaining within $\sim3\cdot10^{-4}$, due to the decrease in the recombination coefficient. Helium remains completely neutral, so $x_{HII}=x_e$. 

In Fig. \ref{hcf_pmf}, we show the main heating/cooling functions for $10<z<300$ for the model with $B_0=0.6$ nGs. Note that the value of the heating function $\Gamma_{pmf}=\Gamma_{mfdt}+\Gamma_{mfad}$ is significantly larger than the value of the heating function due to Compton scattering of the CMB radiation on free electrons, but slightly smaller than the value of the adiabatic cooling function.  
The value of the cooling function $\Lambda_{H_2}$, caused by losses due to excitation of rotational-vibrational lines of the most abundant hydrogen molecule H$_2$ at $z<50$, has significantly increased compared to the standard model (Fig. \ref{hcf}). At $z=20$ , this increase is by 10 orders of magnitude (!), although $\Lambda_{H_2}$ remains, nevertheless, negligibly small compared to the main heating/cooling mechanisms. On the other hand, this indicates that the intensity of emission in molecular lines significantly increases in such models. In the models with $B_0=0.1$ nGs $\Gamma_{pmf}$ becomes larger than $\Gamma_{C_{cmb}}$ at $z<40$. 

The excitation temperature of the upper level of the hyperfine structure of the ground state of hydrogen, or spin temperature, becomes greater than the temperature of the CMB in models with $B_0\gtrsim0.4$ nGs (dotted lines in Fig. \ref{xT_pmf}).

\subsection{Model with self-annihilating dark matter}

\begin{figure*}[htb]
\includegraphics[width=0.495\textwidth]{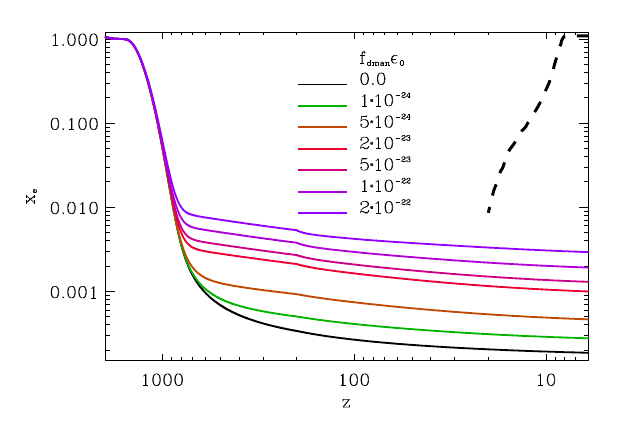}
\includegraphics[width=0.495\textwidth]{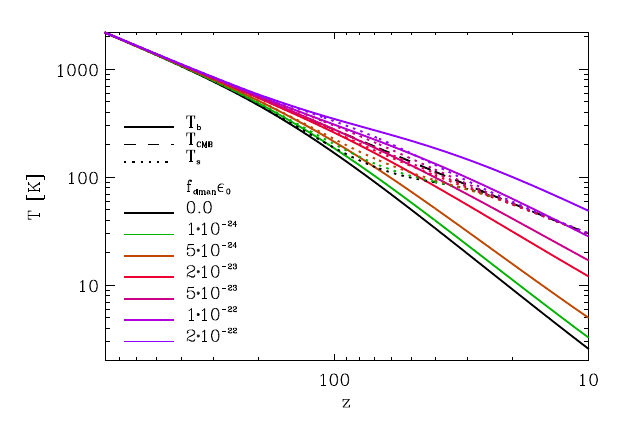}
\caption{Dependence of ionized hydrogen fraction (left) and baryonic matter temperature (right) on redshift in the models with self-annihilating dark matter with different values of the cumulative parameter $f_{dman}\epsilon_0$, which increase for lines from bottom to top. The dashed line in the left panel is the same as in Fig. \ref{xT_pmf}.}
\label{xT_dman}
\end{figure*}
\begin{figure}[htb]
\includegraphics[width=0.5\textwidth]{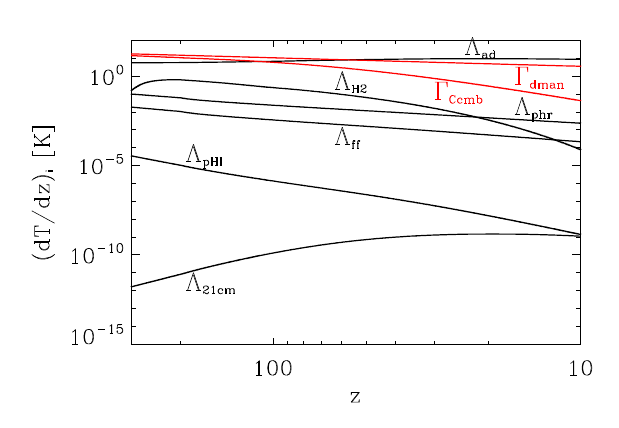} 
\caption{Heating/cooling functions in the model with $f_{dman}\epsilon_0=2\cdot10^{-22}$.}
\label{hcf_dman}
\end{figure}

Standard $\Lambda$CDM model has only one parameter that corresponds to cold dark matter, namely its mass density. There are only upper limits on other admissible parameters, which follow from astrophysical observations and laboratory experiments in which the predicted signals have not been registered under the assumption of a certain nature of dark matter particles. Here we  estimate the values of the parameters of self-annihilating dark matter, at which the change in the ionization and thermal history of baryonic matter in the Dark Ages epoch would manifest itself in the characteristics of the hydrogen 21 cm line. For this, we will use a simple model of self-annihilating dark matter, which is proposed in the work \cite{Chluba2010}. It is assumed that the dark matter particle can self-annihilate with an effective cross-section $\langle\sigma v\rangle$, averaged over the velocities of dark matter particles with mass $m_{dm}$.
The energy that goes to heating baryonic matter per unit volume per unit time is equal to
\begin{eqnarray}
\Gamma_{dman}&=&1.6\cdot10^{-12}f_{dman}g_h \epsilon_0n_H(1+z)^3 \quad \frac{\mbox{\rm erg}}{\mbox{\rm cm$^3$s}},   \\
\epsilon_0&=&4.26\cdot10^{-28}\left[\frac{100\mathrm{GeV}}{m_{dm}}\right]\left[\frac{\Omega_{dm}h^2}{0.12}\right]^2\left[\frac{\langle\sigma v\rangle}{10^{-29}\mbox{\rm cm$^3$/s}}\right]\nonumber
\label{Gan}
\end{eqnarray} 
where $\epsilon_0$ is a dimensionless parameter of the self-annihilation of dark matter particles, $f_{dman}$ is the fraction of the released energy absorbed by baryonic matter, and $g_h=(1 + 2x_{HII} + f_{He}(1 + 2x_{HeII}))/3(1+f_{He})$ is the fraction of that energy that goes to heating the gas.

The products of the annihilation of dark matter particles with an energy of $c^2m_{dm}\gg20$ eV are capable of ionizing hydrogen and helium. Within the framework of the model \cite{Chluba2010}, the ionization rate can be calculated as follows:
\begin{eqnarray*}
(1+z)H\frac{dx_{HI}}{dz}|_{dman}=0.0735f_{dman}g_{ion}^{(HI)}\frac{\epsilon_0n_H(1+z)^3}{n_{HI}(1+f_{He})}, \\
(1+z)H\frac{dx_{HeI}}{dz}=0.04065f_{dman}g_{ion}^{(HeI)}\frac{\epsilon_0n_Hf_{He}(1+z)^3}{n_{HeI}(1+f_{He})},
\end{eqnarray*}
where $g_{ion}^{(HI)}=(1-x_{HII})/3$ and $g_{ion}^{(HeI)}=(1-x_{HeII})/3$ these are the fractions of the released energy that go to the ionization of hydrogen and helium, respectively. As we can see, $g_h\propto g_{ion}\propto 1/3$, which means that approximately one-third of the energy injected into the baryonic component goes to heating, approximately one-third to the ionization of hydrogen and helium, and the rest to the excitation of the energy levels of neutral atoms. Since the phenomenological model \cite{Chluba2010} does not propose a specific model of dark matter particles that self-annihilate, we will estimate the ionization and temperature of the baryonic gas for different values of the cumulative dimensionless parameter $f_{dman}\epsilon_0$. 

The results are shown in Fig. \ref{xT_dman}. Since the annihilation rate of dark matter particles is proportional to the square of their concentration, the contribution to the ionization and thermal history of baryonic matter in the Dark Ages epoch is significant for values of the combined parameter $f_{dman}\epsilon_0\ge10^{-24}$.  The cosmological recombination of hydrogen and helium becomes noticeably different from the standard one in models with $f_{dman}\epsilon_0\gtrsim10^{-22}$. The temperature of the baryonic component at such values is greater than the temperature of the CMB at $200<z<10$. 

The value of the heating function $\Gamma_{dman}$ in the model with $f_{dman}\epsilon_0= 2\cdot10^{-22}$, shown in Fig. \ref{hcf_dman},  is comparable to adiabatic cooling and greater than the value of the heating function by CMB radiation $\Gamma_{C_{cmb}}$ на $z<300$.

\subsection{Model with decaying dark matter}

\begin{figure*}[htb] 
\includegraphics[width=0.495\textwidth]{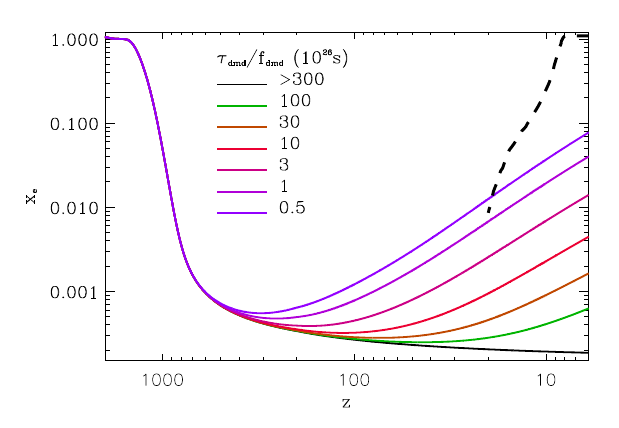}
\includegraphics[width=0.495\textwidth]{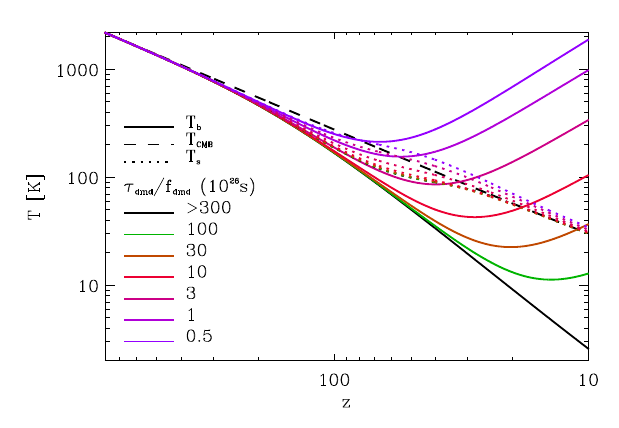}
\caption{Dependence of ionized hydrogen fraction (left) and baryonic matter temperature (right) on redshift in the models with decaying dark matter with different values of the cumulative parameter $\tau_{dmd}/f_{dmd}$, which decrease for lines from bottom to top. The dashed line in the left panel is the same as in Fig. \ref{xT_pmf}.}
\label{xT_dmd}
\end{figure*}
\begin{figure}[htb]
\includegraphics[width=0.5\textwidth]{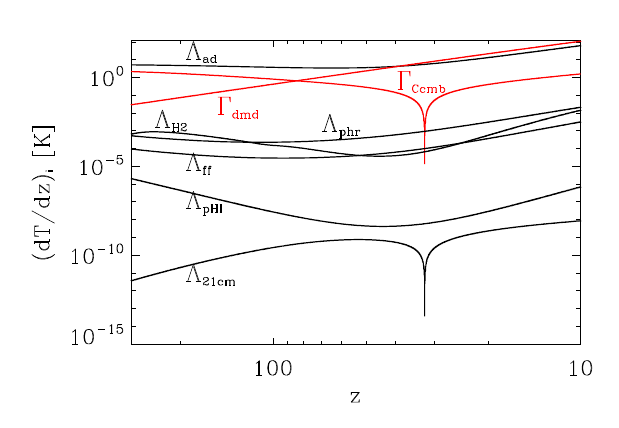} 
\caption{Heating/cooling functions in the model with $\tau_{dmd}/f_{dmd}=3\cdot10^{26}$ c.}
\label{hcf_dmd}
\end{figure}

Dark matter particles can be decaying particles with a lifetime much larger than the age of the Universe. The key parameters of such model are the fraction of dark matter that decays, their characteristic lifetime, and the fractions of particles that go into heating, ionization, and excitation of atoms. Here, as in previous models, we will estimate the values of the decaying dark matter parameters for which the ionization and thermal history of baryonic matter in the Dark Ages epoch differs from the standard one, but does not contradict observational constraints on the optical depth of reionization. To do this, we will use the phenomenological decaying dark matter model \cite{Liu2018}, according to which the heating function can be represented as follows
\begin{eqnarray}
\Gamma_{dmd}=1.69\cdot10^{-8}f_{dmd}g_h\left[\frac{\Omega_{dm}h^2}{0.12}\right]\frac{(1+z)^3}{\tau_{dmd}} \,\, \frac{\mbox{\rm erg}}{\mbox{\rm cm$^3$s}}, 
\end{eqnarray} 
where $\tau_{dmd}$ is the decay lifetime of dark matter particles in seconds, $f_{dmd}$ is the fraction of the released energy that is absorbed by baryonic matter, $g_h$ is the fraction of it that goes to heating the gas. 

Products of the decay of dark matter particles with energies in the keV--GeV range can ionize hydrogen and helium atoms.  
\begin{eqnarray*}
(1+z)H\frac{dx_{HI}}{dz}=7.35\cdot10^{-2}\frac{g_{ion}^{(HI)}}{g_h}\frac{n_H\Gamma_{dmd} }{n_{HI}(1+f_{He})}, \\
(1+z)H\frac{dx_{HeI}}{dz}=4.065\cdot10^{-2}\frac{g_{ion}^{(HeI)}}{g_h}\frac{n_Hf_{He}\Gamma_{dmd}}{n_{HeI}(1+f_{He})},
\end{eqnarray*}
where the fractions of the released energy that go to heating and ionization of hydrogen and helium, $g_h$, $g_{ion}^{(HI)}$ and $g_{ion}^{(HeI)}$, are calculated in the same way as in the previous case of self-annihilating dark matter. 

The results of calculations of the ionization and thermal history of the gas from the epoch of cosmological recombination of hydrogen through the Dark Ages to $z=10$ for values of the total parameter $\tau_{dmd}/f_{dmd}\in5\cdot10^{25} - 10^{28}$s are presented in Fig. \ref{xT_dmd}. For values of this parameter greater than $3\cdot10^{28}$ s, the ionization and thermal history of the gas is practically indistinguishable from the history in the standard $\Lambda$CDM. For the values less than $10^{25}$ s, the ionization of hydrogen at $z\sim20-10$ is higher than the 2$\sigma$ upper limit set in the Planck experiment \cite{Planck2020a} (dashed line in the left panel of Fig. \ref{xT_dmd}). 

The evolution of the gas temperature in this model, presented in the right panel of Fig. \ref{xT_dmd}, is significantly different from the evolution in other models presented here: $T_b\propto T_{min}+T_c(11-z)$. The rapid growth of $x_e$ and $T_b>T_{cmb}$ manifests itself in the change of the nature of the heating function $\Gamma_{C_{cmb}}$ and $\Lambda_{21}$, as can be seen in Fig. \ref{hcf_dmd}. The first has become a cooling function, the second a heating function. This is explained by the fact that $\Gamma_{C_{cmb}}\propto (T_{cmb}-T_b)$ and $\Lambda_{21cm}\propto\left(n_{HI_0}C_{01}-n_{HI_1}C_{10}\right)$ \cite{Seager1999}. 

\subsection{Model with additional cooling}
\begin{figure}[htb] 
\includegraphics[width=0.495\textwidth]{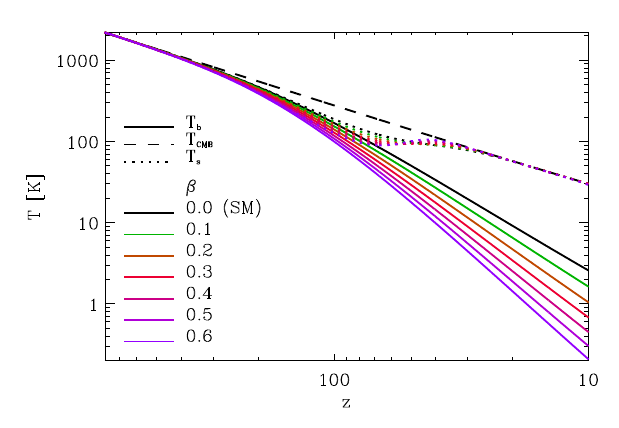}
\caption{Baryonic matter temperature versus redshift in the models with additional cooling with different values of $\beta$ , which increases for lines from top to bottom.}
\label{T_add_cool}
\end{figure}

Among other non-standard dark matter models, those that interact with baryonic matter, causing its cooling, are considered \citep{Munoz2015,Ali2015,Essig2017,Barkana2018}. To estimate the impact of such models on the thermal history of baryonic matter in the Dark Ages and their impact on the formation of the 21 cm absorption line in neutral hydrogen at $z\sim90$, we will consider here a simple cooling model: $\Lambda_{nSM}=\beta\Lambda_{ad}$. The results of calculations of the evolution of the temperature of the baryonic gas in the Dark Ages for $\beta=$0.1, 0.2, ... 0.6 are presented in Fig. \ref{T_add_cool}. Comparison with \cite{Munoz2015} shows that our simple phenomenological model reproduces the course of $T_b(z)$ in a physical model of dark matter that interacts with baryonic matter. This leads to additional cooling of the gas in the Dark Ages.

\section{The 21 cm line from the Dark Ages}

The 21 cm line signal from the Dark Ages can be a source of information about the ionization and thermal history of baryonic matter in this epoch, since the intensity of the line depends on the concentration of neutral hydrogen and the kinetic temperature of the gas. Since these are determined in the standard $\Lambda$CDM model by the residual ionization after cosmological recombination and the competition between adiabatic cooling and heating through Compton scattering of CMB quanta by free electrons, then the free parameters of the solutions of the equations (\ref{Tb})--(\ref{H}) are the baryonic component density $\Omega_b$, the dark matter density $\Omega_{dm}$ and the Hubble constant $H_0$. By measuring such a signal, the values of these parameters can be refined. In non-standard cosmological models with additional heating and ionization, or cooling, as we have seen, the thermal and ionization history may differ significantly from that in the $\Lambda$CDM model, even if the parameters of these models do not go beyond the limits of modern constraints on them. Therefore, the measurement of the 21 cm line signal from the Dark Ages can be an additional test of such cosmological models.

At $z<850$, adiabatic cooling dominates the heating caused by Compton scattering of the CMB radiation on free electrons, and the kinetic temperature of the gas drops faster than the CMB temperature. The spin temperature $T_s$, which reflects the population of the hyperfine structure levels of hydrogen in the Dark Ages epoch, is determined by the processes of excitation and deactivation by CMB photons and collisions with electrons, protons, and neutral hydrogen atoms (see \cite{Novosyadlyj2020} for details), is well described by the relation
\begin{equation}
T_s=T_b\frac{T_{cmb}+T_0}{T_b+T_0}=\frac{(1+x_c)T_{cmb}}{T_b+x_cT_{cmb}},
\label{Tsda}
\end{equation}
which is obtained under the condition of quasi-stationarity of the corresponding kinetic equation. Here, $x_c\equiv T_0/T_{cmb}$ is the collisional coupling parameter, $T_0=h_P\nu_{21}C_{10}/k_B A_{10}$, $h_P$ is the Planck constant, $\nu_{21}$ is the laboratory frequency of the 21 cm line, $A_{10}$ is the Einstein spontaneous transition coefficient, $C_{10}$ is the collisional deactivation rate by electrons, protons, and neutral hydrogen atoms. In Figs. \ref{xeT_Omb}, \ref{xT_pmf}, \ref{xT_dman}, \ref{xT_dmd}, \ref{T_add_cool} the dotted lines show the results of calculations of the spin temperature in the models that we analyze here.

Since the frequency of the 21 cm line $\nu_{21}$ is in the Rayleigh-Jeans range of the energy distribution of the CMB, it is convenient to use the brightness temperature $T_{br}$ instead of the intensity: $I_{\nu}=2k_BT_{br} \nu^2/c^2$. In this case, the useful signal is the difference between the intensities of the CMB and the line emission $\delta I_{\nu}=(I_{\nu}-I_{\nu}^{cmb})/(1+z)$ and the expression for the differential brightness temperature in the 21 cm line, which is obtained by solving the transfer equation of radiation, is as follows \citep{Madau1997,Zaldarriaga2004,Furlanetto2006,Pritchard2012}:
\begin{eqnarray}
\delta T_{br}(z)=23x_{HI}(z)\left[\left(\frac{0.15}{\Omega_m}\right)\left(\frac{1+z}{10}\right)\right]^{\frac{1}{2}}\times\nonumber\\
\hskip-2cm\left[\frac{\Omega_bh}{0.02}\right]\left[1-\frac{T_{cmb}(z)}{T_s(z)}\right] \quad \mbox{\rm mK.}
\label{dTbr}
\end{eqnarray}
As we can see, the differential brightness temperature is proportional to the difference between $T_{s}$ and $T_{cmb}$: if $T_{s}<T_{cmb}$ we will have an absorption line, and if $T_{s}>T_{cmb}$ we will have an emission line. The line disappears when $T_s\rightarrow T_{cmb}$.

\subsection{Standard $\Lambda$CDM model}

\begin{figure*}[htb] 
\includegraphics[width=0.495\textwidth]{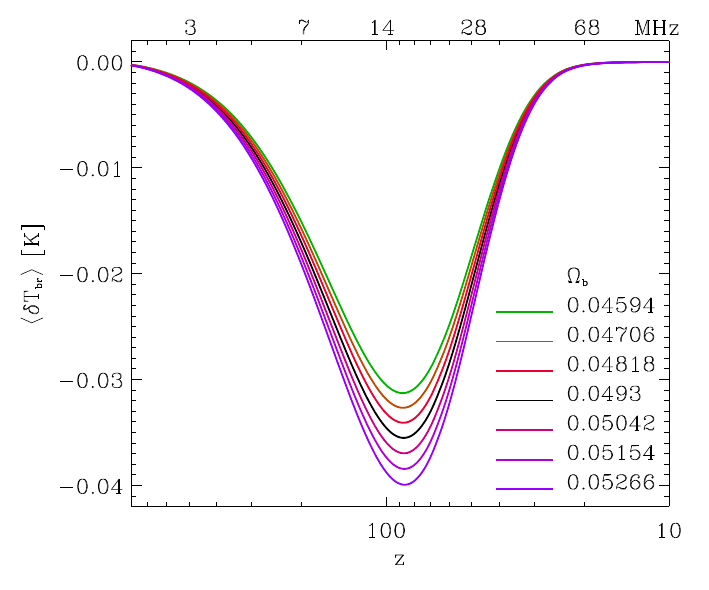}
\includegraphics[width=0.495\textwidth]{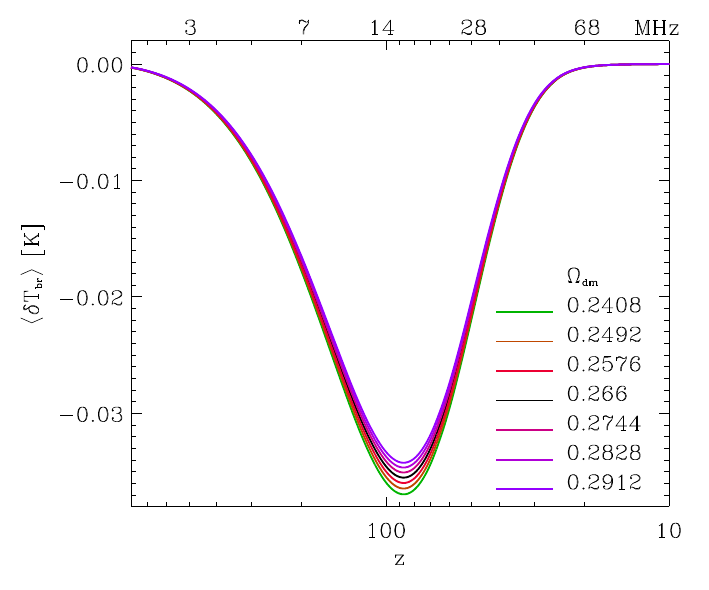}
\caption{Absorption line of 21 cm neutral hydrogen from the Dark Ages in the standard $\Lambda$CDM model with different values of $\Omega_b=0.0493\pm1\sigma,\,\pm2\sigma,\,\pm3\sigma$ (left) and $ \Omega_{dm}=0.266\pm1\sigma,\,\pm2\sigma,\,\pm3\sigma$ (right).}
\label{H21_Ombd}
\end{figure*}
\begin{figure}[htb]
\includegraphics[width=0.5\textwidth]{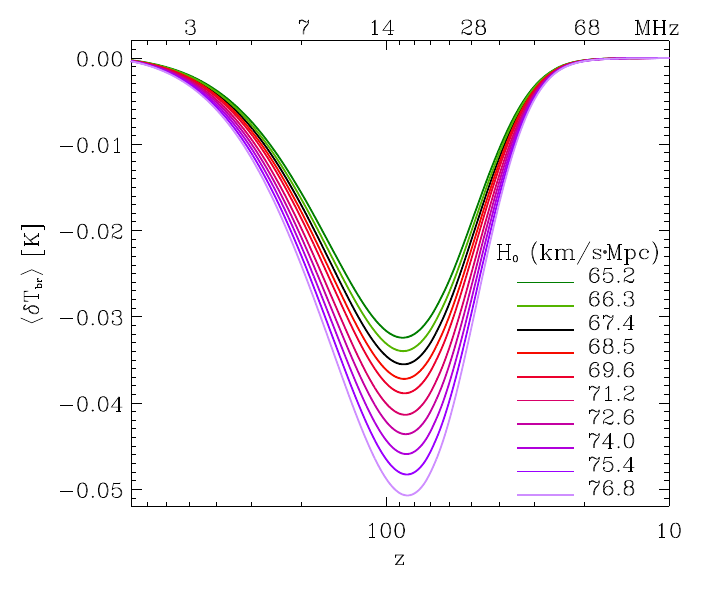} 
\caption{The absorption line of 21 cm neutral hydrogen from the Dark Ages in the standard $\Lambda$CDM model with different values of $H_0=67.36\pm1\sigma,\,\pm2\sigma$ km/c$\cdot$Mpc and $H_0=74 \pm1\sigma,\,\pm2\sigma$ km/c$\cdot$Mpc.}
\label{H21_H0}
\end{figure}

To assess the possibilities of refining the values of the cosmological parameters of the standard  $\Lambda$CDM model based on measurements of the differential brightness temperature from the Dark Ages epoch, we calculated it for different values of $\Omega_b$ and $\Omega_{dm}$ within the 3$\sigma$ limits, provided that equation (\ref{H}) is satisfied for the current epoch,
\begin{equation}
\Omega_b+\Omega_{dm}+\Omega_K+\Omega_\Lambda=1.
\label{Fr1}
\end{equation}
We also calculate it for different values of $H_0$ in the vicinity of the values obtained on  the base of measurements of the power spectrum of temperature and polarization fluctuations of the CMB \citep{Planck2020a} and the magnitude-redshift test for type Ia supernovae \citep{Riess2022}. The results of the calculations are presented in Figs. \ref{H21_Ombd} and \ref{H21_H0}. The dependence of the line position, depth, and effective width on these parameters, in addition to the obvious one from the relation (\ref{dTbr}), occurs through the dependence of $x_{HI}$ on them. Since the state of baryonic matter is practically insensitive to variations in $\Omega_K$ and $\Omega_\Lambda$, then to fulfill (\ref{Fr1}) when varying $\Omega_b$ and $\Omega_{dm}$ relative to the values of the reference model, we will vary $\delta\Omega_\Lambda=-\delta\Omega_b$ and $\delta\Omega_\Lambda=-\delta\Omega_{dm}$, respectively.

As seen in Figs. \ref{H21_Ombd} and \ref{H21_H0}, variations in $\Omega_b$, $\Omega_{dm}$ and $H_0$ change the depth and width of the $z$-profile of the line. Increasing $\Omega_b$ increases the depth of the absorption line (decreases $\delta T_{br}^{min}$) due to the higher density of absorbers of CMB quanta and the greater efficiency of collisional processes in deactivating the upper level of the hyperfine structure of the hydrogen atom, which brings the spin temperature $T_s$ closer to $T_b$. Conversely, increasing $\Omega_{dm}$ decreases the depth of the line due to the increase in the expansion rate of the Universe in the Dark Ages epoch, $Н(z)\sim\sqrt{(\Omega_b+\Omega_{dm})(1+z)^3}$. The variation of the $z$-profile of the line due to the variation of the Hubble constant $H_0$ is shown in Fig. \ref{H21_H0}: increasing $H_0$ increases the depth of the line, since the concentration of neutral hydrogen atoms $\propto H_0^2$. In all cases, the minima are found in the range of redshifts $86\le z\le88$ ($\sim16.1\pm0.2$) MHz. The 2$\sigma$ deviations of these parameters from the reference values are 4.5\% for $\Omega_b$, 5.2\% for $\Omega_{dm}$ and 1.6\% for $H_0$. The variations in the depth of the 21 cm absorption line for them are 8.2\%, 2.5\% and 4.3\%, respectively. The full width at half maximum of the line for the models in Figs. \ref{H21_Ombd} and \ref{H21_H0} is in the range of 23.2---23.6 MHz. 

\subsection{Non-standard cosmological models}

\begin{figure*}[htb] 
\includegraphics[width=0.495\textwidth]{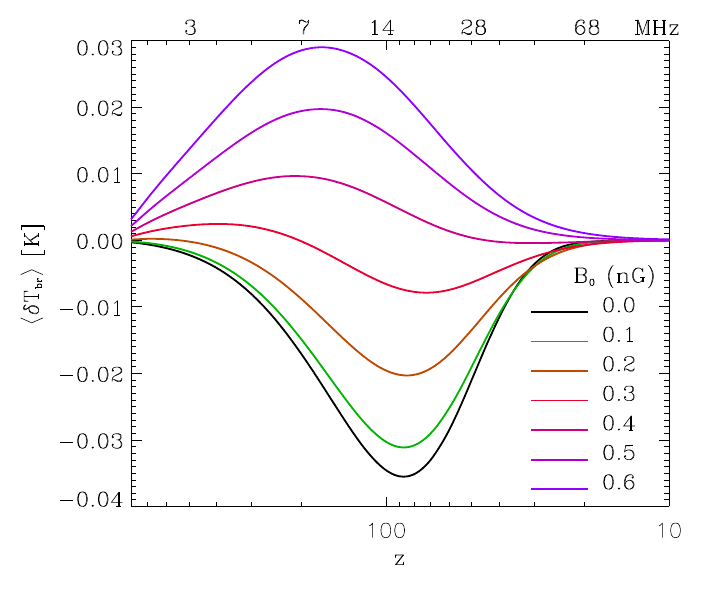}
\includegraphics[width=0.495\textwidth]{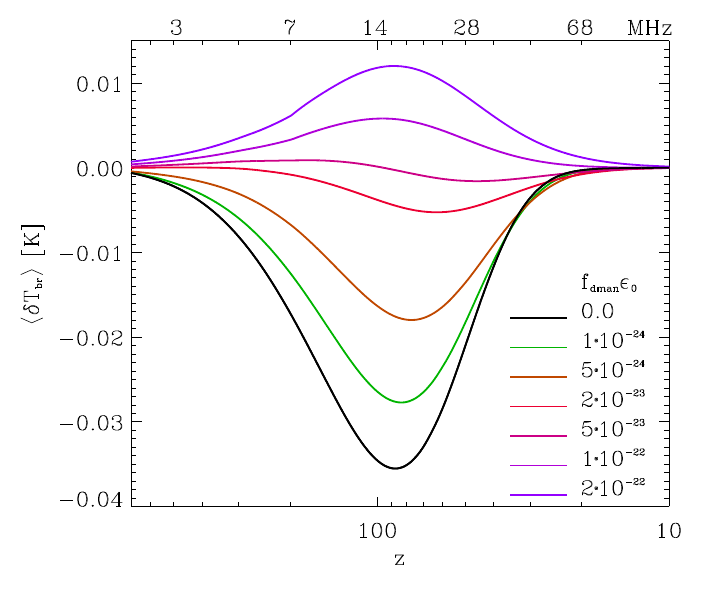}
\includegraphics[width=0.495\textwidth]{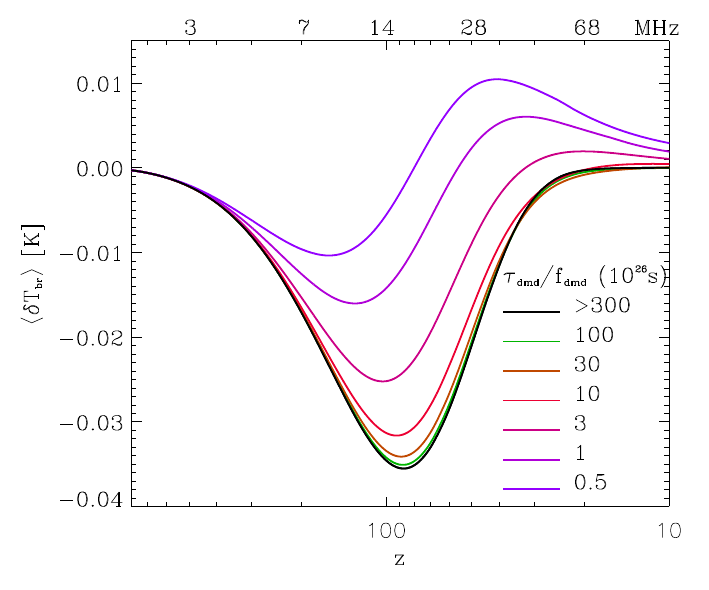}
\includegraphics[width=0.495\textwidth]{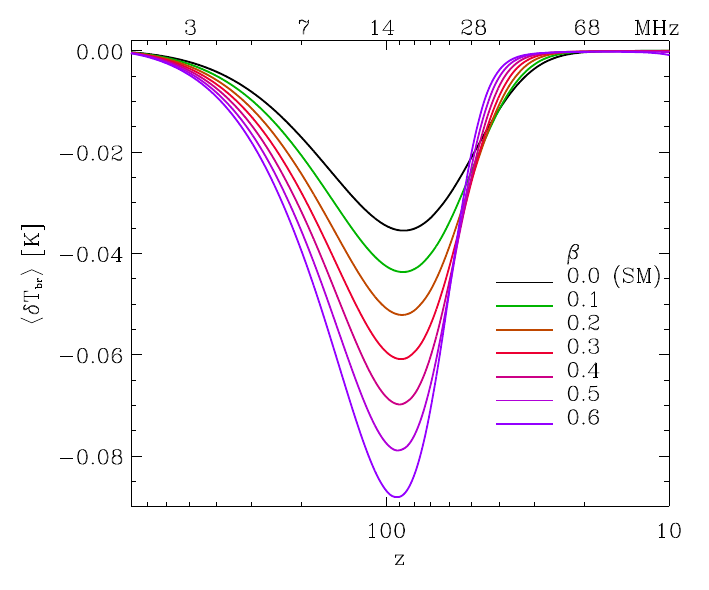}
\caption{Absorption line 21 cm neutral hydrogen from the Dark Ages in non-standard models with a primordial magnetic fields (top left), self-annihilating dark matter (top right), with decaying dark matter (bottom left), and in a model with additional cooling. Model parameters are the same as in the corresponding figures of the previous paragraph.}
\label{H21_nSM}
\end{figure*}

The results of the calculations presented in the previous paragraph demonstrate a significant change in the ionization of hydrogen and the temperature of the baryonic gas in the Dark Ages epoch in non-standard cosmological models with a primordial magnetic field, with self-annihilating dark matter, with decaying dark matter, and in a model with additional cooling. Since the ionization and heating by the light of the first sources in the Cosmic Dawn and Reionization epoch erases the previous ionization and thermal history, the signal in the 21 cm line of neutral hydrogen may be the only source of information about such models at sub-threshold values of the signals of other evidence, such as the registration of products of self-annihilation, decay, other direct or indirect manifestations.

Fig. \ref{H21_nSM} shows the results of calculations of the differential brightness temperature in the 21 cm line of neutral hydrogen from the Dark Ages epoch ($10\le z\le800$, $1.8\le \nu_{obs}\le130$ MHz) for all non-standard models for which the ionization and thermal history has been calculated in the previous paragraph. As can be seen, in cosmological models with additional heating or cooling, the profile of the 21 cm line of neutral hydrogen differs significantly from the profile in the standard $\Lambda$CDM.

\subsubsection{Primordial magnetic field}

As $B_0$ increases, the depth of the absorption line decreases, disappears, and turns into an emission line (left panel in the upper row of Fig. \ref{H21_nSM}). The transition occurs at $B_0\approx0.3$ nGs by the appearance of a small emission "hump" $\delta T_{br}\approx2$ mK in the low-frequency wing of the absorption line at $z\approx390$, which corresponds to $\nu_{obs}\approx3.6$ MHz. Its width at half maximum is $\Delta\nu_{obs}\approx3.8$ MHz. The position of the absorption line in this magnetic field model shifted towards higher frequencies $\nu_{obs}\approx20$ MHz ($z\approx72$), the depth decreased to $\approx-8$ mK, the width at half maximum  $\Delta\nu_{obs}\approx28$ MHz. At larger values of $B_0$ the absorption line disappears in the high-frequency wing, and the emission line shifts towards higher frequencies, its amplitude and width increase. At $B_0=0.6$ nGs, its maximum $\delta T_{br}\approx30$ mK is at a frequency of $8.3$ MHz ($z\approx170$), the width at half maximum $\Delta\nu_{obs}\approx16$ MHz.

\subsubsection{Self-annihilating dark matter}

The behavior of the 21 cm line profile in models with self-annihilating dark matter (right panel in the top row of Fig. \ref{H21_nSM}) with increasing values of the total parameter $f_{dman}\epsilon_0$ is qualitatively similar to the behavior in models with primordial magnetic fields: the appearance of a "hump" in the low-frequency wing of the line, the disappearance of absorption in the high-frequency wing. In this case, the emission "hump" becomes noticeable at $f_{dman}\epsilon_0\approx5\cdot10^{-23}$: its amplitude $\delta T_{br}\approx2.5$ mK at $\nu_{obs}\approx11$ MHz ($z=128$), half-width at half-maximum $\Delta\nu_{obs}\approx21$ MHz.
At larger values of $f_{dman}\epsilon_0$
, the absorption line disappears in the high-frequency wing, and the emission line shifts to higher frequencies, its amplitude and width increase. At $f_{dman}\epsilon_0\approx2\cdot10^{-22}$, its amplitude $\delta T_{br}\approx13$ mK is at a frequency of $16$ MHz ($z\approx88$), half-width at half-maximum $\Delta\nu_{obs}\approx34$ MHz. 

Larger values of the cumulative parameter, as shown in Fig. \ref{xT_dman}, are constrained by data on the polarization of the CMB at low spherical harmonics, obtained in the Planck experiment \citep{Planck2020a}. Although the qualitative behavior of the profile in the model with self-annihilating dark matter is the same as in the model with an initial magnetic field, they are distinguishable in the quantitative characteristics of the 21 cm line profiles at the upper allowed values of $B_0$ and $f_{dman}\epsilon_0$.

\subsubsection{Decaying dark matter}

In the models with decaying dark matter particles, the ionization and thermal histories are significantly different from those in other models discussed in this work. It is expected that the 21 cm line profiles from the Dark Ages will also be special, as illustrated by the left panel in the bottom row of Fig. \ref{H21_nSM}. The emission "hump" in this model appears in the high-frequency wing of the 21 cm line at $z\approx25$ ($\nu_{obs}\approx55$ MHz), when $\tau_{dmd}/f_{dmd}\approx3\cdot10^{26}$ s.  Its amplitude $\delta T_{br}\approx2$ mK, full width at half maximum $\Delta\nu_{obs}\approx8$ MHz. A deep absorption minimum in this model $\delta T_{br}\approx-25$ mK is in the low-frequency wing of the line at $\nu_{obs}\approx13.7$ MHz ($z\approx103$). With a decrease of the decay lifetime or an increase of the fraction of decaying dark matter particles, that is, with a decrease in the cumulative parameter $\tau_{dmd}/f_{dmd}$ to $5\cdot10^{25}$ s, the amplitude of the emission "hump" increases to 10 mK, shifts towards lower frequencies to 34 MHz ($z\approx41$), its full width at half maximum $\Delta\nu_{obs}$ is $\approx35$ MHz. The absorption minimum in the low-frequency wing of the line decreases to -10 mK, shifts to $\nu_{obs}\approx9$ MHz, and the full width at half maximum decreases to 10 MHz.

The difference in the 21 cm line profile in this model from the profile in previous models is due to the rate of energy release in the corresponding processes: $\propto(1+z)^4$ in the case of primordial magnetic field, $\propto(1+z)^6$ in the case of self-annihilating dark matter and $\propto(1+z)^3$ in the case of decaying dark matter, which describe the corresponding heating functions presented in Section II. 

\subsubsection{Additional cooling}

The 21 cm absorption line from the Dark Ages can also be an effective test of models with additional cooling. Such cooling can be provided by electron scattering of millicharged dark matter particles \citep{Munoz2015,Ali2015,Essig2017}. Renan Barkana \cite{Barkana2018} used such a model to interpret the anomalously deep minimum of the 21 cm absorption line at $z\approx17$, detected by the EDGES team \cite{Bowman2018}. The thermal history of the baryonic gas in a simple phenomenological model with different amounts of additional cooling is shown in Fig. \ref{T_add_cool}. The 21 cm line profiles from the Dark Ages for these models are shown in the right panel of the bottom row of Fig. \ref{H21_nSM}. As expected, with increasing the additional cooling function, which is proportional to the adiabatic cooling of the gas due to the expansion of the Universe $\Lambda_{nSM}=\beta\Lambda_{ad}$, from $\beta=$0 to 0.6, the depth of the absorption line increases from -36 mK to -88 mK. Its position shifts from $z\approx87$ ($\nu_{obs}\approx16$ MHz) to $z\approx92$ ($\nu_{obs}\approx15$ MHz), and the half-width at half-maximum decreases from 23.3 MHz to 16.6 MHz. The depression in the short-wavelength wing of the line is caused by the rapid loss of efficiency of collisional excitation of the hyperfine level of hydrogen due to cosmological expansion (dotted line in Fig. \ref{T_add_cool}). 

These results can be compared to the analogous results in \cite{Barkana2018}. The solid red line in Fig. 2 of the paper \cite{Barkana2018} approximately corresponds to the most probable observed value of the peak absorption in the Cosmic Dawn epoch, registered by EDGES \cite{Bowman2018}. It indicates a minimum absorption in the Dark Ages epoch $\approx-67$  mK at a frequency of $\approx17$ MHz. In our model, this corresponds to $\beta\approx0.4$.  

The results presented in Figs. \ref{H21_Ombd}--\ref{H21_nSM} show that the 21 cm line of the hyperfine structure of the ground state of neutral hydrogen, which is formed in the Dark Ages epoch at $z\sim80-100$, can be an effective test of cosmological models. The question of its detection possibilities remains open. 

\section{The possibilities of detecting the 21 cm line from the Dark Ages}
 
The above discussion shows that radio telescopes in the decameter wavelength range, $\nu = 3-30\,\, \mbox{MHz}\,\,(\lambda = 10 - 100\,\, \mbox{m}$), are necessary for detecting the global signal in the 21 cm line of neutral hydrogen from the Dark Ages. The world's largest operating radio telescope in this frequency range is located in Ukraine, the Ukrainian T-shaped radio telescope UTR-2\footnote{Operated until February 24, 2022, destroyed by Russian forces during the liberation of Kharkiv region in the fall of the same year. Restoration work is currently underway on it.}. It was created by specialists from the Institute of Radio Astronomy of the National Academy of Sciences of Ukraine near the city of Chuhuiv, Kharkiv region. Its team has over half a century of experience in observing, processing, and interpreting radiation from celestial objects of various natures in the decameter wavelength range, which is difficult to observe from Earth due to the high level of interference from both galactic and ionospheric and artificial sources. In this section, we present a brief overview of the analysis of the possibilities of detecting the 21 cm line from the Dark Ages based on the technical characteristics of UTR-2 and the experience of the scientific school of the Institute of Radio Astronomy. A full presentation of the analysis of this complex problem can be found in the accompanying paper \cite{Konovalenko2023}. 

To formulate the requirements for the experimental search for the redshifted 21 cm hydrogen line, it is necessary to carefully consider the features of radio astronomy at decameter waves, both from the point of view of astrophysics and hardware-methodological aspects
\begin{equation}
 \Delta S_{min}= \frac{2k_B T_{noise}}{A_{ef}\sqrt{\Delta f\Delta t} }\left(1+\frac{T_N}{T_a}\right)\,\, \mbox{Jy}, \label{dSmin2}
\end{equation} 
where $\Delta S_{min}$ is the minimum flux density of electromagnetic radiation detected from a radio source,  $T_{noise}$ is the system noise temperature, $A_{ef}$ is the effective area of the radio telescope, $\Delta f$ is the frequency band of registration (frequency resolution), $\Delta t$ is the integration or accumulation time (time resolution), $T_N$ is the noise temperature of the receiver, $T_{a}$ is the antenna temperature (power of the radio signal from the galactic background) at the output of the antenna or antenna element. The latter is calculated as follows: $T_{a}=\eta_{A}T_{B}$, where $\eta_{A}$ is the frequency-dependent signal attenuation, and $T_B$ is the brightness temperature of the foreground. At decameter wavelengths, the system noise temperature is determined mainly by the brightness temperature of the galactic background, $T_{noise} \approx T_{B}(\nu)$, which reaches tens and hundreds of thousands of K depending on the frequency. In the frequency range that we are interested in, this is synchrotron radiation from the interstellar medium with $T_{B}(\nu)\propto\nu^{-2.6}$.
 
The relationship (\ref{dSmin2}) is valid for point radio sources with angular sizes smaller than the width of the radio telescope beam pattern. For cases of extended or isotropic radio sources, such as the global 21 cm hydrogen line signal from the Dark Ages, the expression  can be used for the minimum brightness temperature that can be detected in the frequency band $\Delta f$ over an integration time $\Delta t$
\begin{equation}
\Delta T_{min}=\frac{T_{B}}{\sqrt{\Delta f \Delta t}}\left(1+\frac{T_N}{T_a}\right)\,\, \mbox{Jy}. \label{dTmin}
\end{equation} 
Another characteristic of radio telescopes is the sensitivity reduction factor $m\equiv T_a/(T_a+T_N)$, which is an important component of any experiment. For Ukrainian telescopes, it reaches  $\approx$0.9.
Expression (\ref{dTmin}) can be used to estimate the required signal accumulation time to obtain the minimum brightness temperature $T_{min}$ necessary to detect a signal: $\Delta T_{min}\approx\delta T_{br}/(S/N)$, where the signal-to-noise ratio is taken to be $S/N=10$. Based on the map of galactic background radio emission at a frequency of 20 MHz, given in \cite{Konovalenko2023} (Fig. 9), we can estimate the temperature of the "coldest" regions of the sky in the galactic poles at a given frequency: $T_B(\nu)\approx 2\cdot10^4(\nu/20\,\mbox{MHz})^{-2.6}$ K. For an absorption line with a brightness temperature of $\delta T_{br}=35$ mK at a frequency of $\nu_{obs}=16$ MHz, where $T_B\approx 36000$ K, with observations in a frequency band of $\Delta f=\Delta\nu=25$ MHz in the coldest regions of the sky in the galactic poles, we obtain $\approx60$ days. This optimistic result of the estimate provides grounds for a deeper analysis of the possibilities of detecting the 21 cm line of neutral hydrogen from the Dark Ages.  

Based on radiospectroscopic requirements and the experience of registering recombination lines with the UTR-2 radio telescope at decameter waves \cite{Konovalenko2023}, for a line with the given parameters, the following requirements can be formulated for its detection:
analysis bandwidth $\Delta F=(2-10) \Delta \nu_{L} \approx 50-250$ MHz;
frequency resolution $\Delta f = \Delta \nu_{obs}/(1-10) \approx 25-2.5$ MHz;
number of frequency channels $M=\Delta F/\Delta f_{min} = 20-100$;
sampling frequency $F\geq 2\Delta F= 100-500$ MHz;
sampling resolution $q=16$ bit;
internal resolution from the point of view of radio interference and recombination lines $\Delta f_{int} = 1-10$ kHz;
number of internal channels $M_{int} = 5000-25000$;
angular resolution $\Theta_{A}= 30^\circ - 180^\circ$; 
polarization $N_{p} = 2$;
relative sensitivity $\Delta T_{L}/T_{B} \approx 10^{-6}$;
signal-to-noise ratio $S/N \rightarrow 10$;
fluctuation level on spectra $\tau\leq (\Delta T_{L}/T_{B})/(S/N) = 10^{-7}$;
time resolution is not required (the effect is stationary);
type of spectrometer -- a) digital autocorrelation; b) digital with direct fast Fourier transform.  

The radio telescope GURT of the Institute of Radio Astronomy of NASU meets these requirements, with a bandwidth of 8-80 MHz.

Let us comment and pay attention to some important requirements for an experiment in the field of low-frequency radio spectroscopy. The analysis band should be several times wider than the line width. This ensures the implementation of relative spectroscopic measurements, which are much more accurate than absolute measurements. The intensity in the line is compared to the neighboring level of the spectrum, where there are no lines known to be present. Estimates show that it is practically impossible to achieve a significantly wider band while maintaining maximum sensitivity $m \sim  0.9$ and interference resistance\footnote{Antenna elements on the far side of the Moon will improve the situation in the future.}. The resolution should be several times better than the line width in order to clarify the line profile (the number of channels is several tens). In the presence of narrow-band radio interference, which are present in the decameter wavelength range, the resolution should be quite high $(\leq 10\,\, \mbox{kHz})$, and the number of channels should reach several thousand. The contradiction between the last two requirements is resolved by using higher resolution at the observation stage, and during the secondary processing of the obtained spectra with interference and recombination lines, the latter are removed by means of special digital filtering, and then the spectra themselves are "smoothed" in frequency and according to the expected width of cosmological lines.

A detailed description of natural and instrumental interference in the registration of extragalactic radio emission in the decameter wavelength range and methods for their elimination are presented in the accompanying article \cite{Konovalenko2023}. The methodology of observations at the GURT radio telescope of the National Academy of Sciences of Ukraine for the registration of the 21 cm line from the Dark Ages is also described there. 

\section{Conclusions}
In this work, it is shown that the global signal in the 21 cm line of the ground state of neutral hydrogen from the Dark Ages ($z\sim100$), shifted to the decameter wavelength range by the expansion of the Universe, can be an effective test of some parameters of the standard $\Lambda$CDM model and cosmological models with additional heating/cooling mechanisms. To do this, the evolution of the global (averaged over the sky) differential brightness temperature in this line was calculated in standard and non-standard cosmological models with different parameters.

The standard $\Lambda$CDM model with post-Planck parameters predicts a value of the differential brightness temperature at the center of the absorption line of $\delta T_{br}\approx-35$ mK at $z\approx87$. The observed frequency of the line at the absorption maximum is $\approx$16 MHz, and the effective width of the line is $\approx$25 MHz. The depth of the line is moderately sensitive to $\Omega_b$ and $H_0$, weakly sensitive to $\Omega_{dm}$ and insensitive to other parameters of the standard $\Lambda$CDM model.

The presence of primordial magnetic fields in intergalactic space follows from observations of the emission of blazars in the TeV and GeV energy ranges, which give a lower bound on the value of the induction of such a field: $B_0\simeq10^{-11}$ nGs \cite{Neronov2010,Takahashi2013}. The upper 95\% limit was obtained from complex measurements of temperature fluctuations and polarization of the CMB by the Planck satellite: $B_0\simeq0.9$ nGs. We analyzed the impact of such fields on the thermal history of the Universe in the Dark Ages due to the decay of turbulence and ambipolar diffusion and showed that fields with $B_0\ge0.1$ nGs already significantly reduce the depth of the 21 cm absorption line of the standard $\Lambda$CDM model. At $B_0\sim0.3$ nGs, the absorption line turns into emission in the low-frequency wing of the absorption line. Its amplitude grows $\propto B_0^{3\div4}$. In the studied models, the amplitude of the absorption-emission line is in the range -31--29 mK, the position at the frequency 16--8 MHz, the width at half-maximum 23--21 MHz.
Thus, the global signal in the 21 cm line can indicate the values of $B_0$ of primordial magnetic fields in the range 0.1--0.9 nGs, or lower the upper limit on them.

Natural extension of the $\Lambda$CDM model is the assumption of the diversity of physical properties of dark matter particles. Among them can be both those that self-annihilate, decay and participate in significantly weakened Coulomb interaction with electrons or protons.

Energy injected into the baryonic component according to the model with self-annihilating dark matter is used to heat, ionize, and excite hydrogen and helium atoms. Therefore, the value of the cumulative parameter\footnote{The averaged product of the effective annihilation cross-section and particle velocity, multiplied by the fraction of the density of self-annihilating dark matter from the total dark matter, divided by the particle mass (see expression \ref{Gan}).} of such dark matter model is limited by data on cosmological recombination and the optical depth of reionization. In the range of its values $10^{-24}\le f_{dman}\epsilon_0\le 2\cdot10^{-22}$ , such dark matter can be detected by its effect on the parameters of the 21 cm line from the Dark Ages epoch: the depth/height and width of the absorption/emission dip/peak. In the studied models, the amplitude of the absorption-emission line is in the range -35--12 mK, the position at the frequency of 18--16 MHz, the width at half maximum of 24--34 MHz.

A model with decaying dark matter, the decay products of which heat and ionize the baryonic component, affects the 21 cm line profile at values of the cumulative parameter\footnote{The ratio of the duration of the decay lifetime to the fraction of decaying dark matter} $\tau_{dmd}/f_{dmd}<3\cdot10^{28}$ s. At its values $<3\cdot10^{26}$ s, the ionization becomes greater than the 2$\sigma$ limit of reionization at $z\sim20$, obtained by the Planck collaboration \cite{Planck2020a}. The model is distinguished by the pecular asymmetry of the line profile: absorption in the low-frequency wing of the line and emission in the high-frequency wing. In the studied models, the amplitude of the absorption line is in the range -10 -- -35 mK, the position at the frequency of 9--16 MHz, the width at half maximum is 10--23 MHz. The amplitude of the emission line is in the range 2--10 mK, the position at the frequency of 34--55 MHz, the width at half maximum is 8--35 MHz.

A model with additional cooling predicts a deeper line depth the greater the value of the cooling function at $z\sim100$. If it is 60\% of the adiabatic cooling, then the depth of the absorption line is -88 mK at a frequency of 15.3 MHz. In our model, which is close to the most optimal from work \cite{Barkana2018}, which explains the absorption line at a frequency of $\approx78$ MHz \cite{Bowman2018} by scattering of electrons on millicharged dark matter particles, the depth of the line is $\approx70$ mK at a frequency of $\approx16$ MHz.

An analysis of the capabilities of detecting a global 21 cm line signal from the Dark Ages using domestic radio telescopes in the decameter wavelength range, UTR-2 and GURT, proves the prospects for setting and implementing such an observational task.

\section*{Acknowledgements}
This work was carried out within the framework of the project of the Ministry of Education and Science of Ukraine "Modeling the luminosity of elements of the large-scale structure of the early Universe and the remnants of galactic supernova stars and observations of variable stars" (state registration number 0122U001834) and with the support of the International Center for Future Science and the College of Physics of Jilin University (China). The work on it was possible thanks to the resilience and courage of the Armed Forces of Ukraine, who are resisting the unprovoked aggression of Russia against Ukraine.

\end{document}